\documentclass[journal]{IEEEtran}
\usepackage{subfigure}
\usepackage{cite}
\usepackage{hyperref}
\usepackage{csquotes}
\usepackage{balance}
\usepackage{color}
\usepackage{subfigure}
\usepackage{graphicx}
\usepackage{amsmath}
\usepackage{lipsum}

\usepackage{subfigure}
\usepackage{cite}
\usepackage{hyperref}
\usepackage{csquotes}
\usepackage{balance}
\usepackage{color}
\usepackage{subfigure}
\usepackage{graphicx}
\usepackage{tabularx}
\usepackage{lipsum}
\usepackage{indentfirst}
\usepackage{csquotes}
\usepackage{soul}
\MakeOuterQuote{"}
\begin{document}
\title{Pathway to Secure and Trustworthy ZSM for LLMs: Attacks, Defense, and Opportunities}
\author{Sunder Ali Khowaja \IEEEmembership{Senior Member, IEEE}, Parus Khuwaja, Kapal Dev \IEEEmembership{Senior Member IEEE}, Hussam Al Hamadi, and Engin Zeydan \IEEEmembership{Senior Member IEEE} 
\thanks{Sunder Ali Khowaja is with School of Computing, Dublin City University, and CONNECT Centre, Dublin, D09 V209, Ireland. Email: sunderali.khowaja@dcu.ie}
\thanks{Parus Khowaja is with Institute of Business Administration, University of Sindh, Jamshoro. (e-mail:Parus.khuwaja@usindh.edu.pk).}
\thanks{Kapal Dev is associated with CONNECT Centre and Department of Computer Science and  Munster Technological University, Bishopstown, Cork, T12 P928, Ireland,e-mail: (kapal.dev@ieee.org)}
\thanks{Hussam Al Hamadi with College of Engineering and IT 
University of Dubai, e-mail: (Halhammadi@ud.ac.ae)}
\thanks{Engin Zeydan with Centre Tecnològic de Telecomunicacions de Catalunya (CTTC), Barcelona, Spain, 08860, e-mail: (ezeydan@cttc.es)}
}

%


\maketitle

\begin{abstract}

Recently, large language models (LLMs) have been gaining a lot of interest due to their adaptability and extensibility in emerging applications, including communication networks. It is anticipated that zero touch network and service management (ZSM) will be able to support mobile edge computing networks and LLMs as a service, as they aim to provide network automation and service management without requiring manual intervention. However, LLMs are vulnerable to data and model privacy issues that affect the trustworthiness of LLMs to be deployed for user-based services. In this paper, we explore the security vulnerabilities associated with fine-tuning LLMs in ZSM in particular the membership inference attack. We define the characteristics of an attack network that can perform a membership inference attack if the attacker has access to the fine-tuned model for the downstream task. We show that the membership inference attacks are effective for any downstream task, which can lead to a personal data breach when using LLM as a service. The experimental results show that the attack success rate of maximum 92\% can be achieved on named entity recognition task. Based on the experimental analysis, we discuss possible defense mechanisms and present possible research directions to make the LLMs more trustworthy in the context of 6G networks. 
\end{abstract} 

\section{Introduction}\label{sec:intro}

The emergence of attention networks has been a stepping stone for transformer architectures, which also led to the introduction of large language models (LLMs). More recently, LLMs are seen as the most significant advance in the field of artificial intelligence (AI) and a potential pathway to artificial general intelligence (AGI) \cite{surveyLLM}. Every tech giant is in a race to advance in the field of LLMs by leveraging generative AI (GAI). Notable examples of LLMs from the tech giants are GPT-4 from OpenAI, LLaMA-3 from Meta and PALM from Google. However, there are also new players in this field that surpass the performance of the LLMs mentioned above. These include Mistral (in collaboration with NVIDIA), DCLM from Apple, xLAM from Salesforce, v2 chat from Deepseek, Groq, Claude, SmolLM and many more. These LLMs are trained on diverse and large amounts of datasets scraped or curated from the Internet. Some LLMs focus on increasing model size, such as GPT, while others find new ways to improve the generalization of LLMs through data curation, model quantization, and innovative techniques. Examples of such LLMs are Claude, LLaMA, DCLM and Groq, which have recently outperformed GPT on various language tasks. In continuation of the above-mentioned advances in LLMs, several enterprises are leveraging pre-trained encoders of LLMs to varying degrees to develop their own customized solutions for various applications and sectors, including healthcare, education, law and industrial automation. In view of the rapid development of LLMs, it can be assumed that LLMs will soon also be deployed on edge and handheld devices. Several studies have indicated that the current iteration of networks will not be able to support a plethora of services offered by LLMs. Therefore, researchers are working intensively on the next iteration of communication systems, i.e. Sixth generation (6G) and zero touch network and service management (ZSM), to meet the above requirements \cite{Net4AI}. Furthermore, as AI is an integral part of ZSM, it is assumed that LLMs will be used intrinsically to optimize resources and performance while enabling human-centric customized services to users. \\

\begin{figure*}[h]
\centering
  \includegraphics[width=\linewidth]{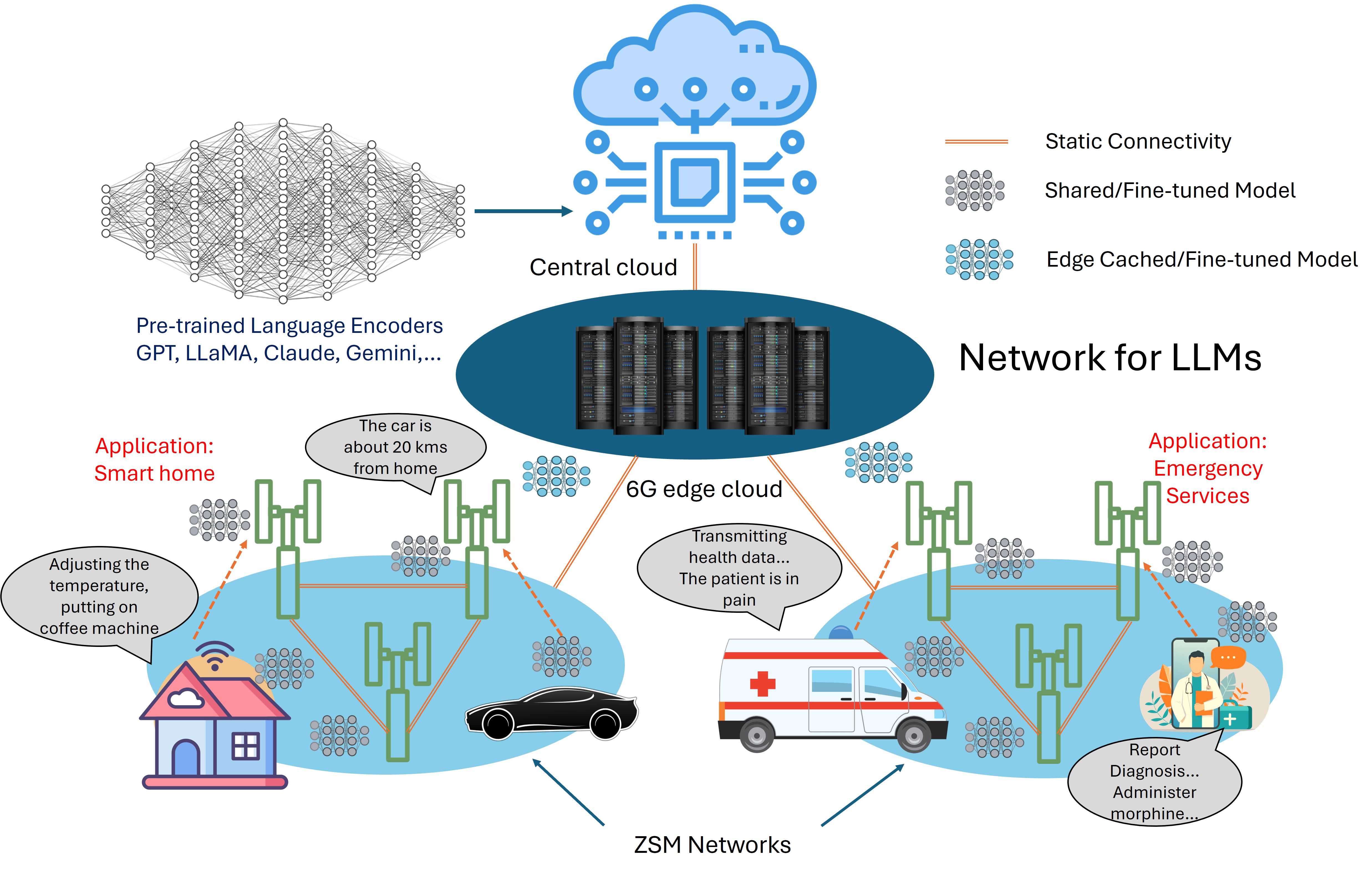}
  \caption{Network for LLMs illustration in ZSM with examples of smart homes and emergency services. The LLMs can be used by the central cloud and shared with the 6G Edge Cloud. The 6G edge cloud then share either the parameters to the ZSM network or users to fine-tune the network for personalization, or it shares the cached version of the model to provide a specific service.}
  \label{Fig1}
\end{figure*}

\indent The standardization process towards 6G systems and ZSM networks is already progressing steadily. It is assumed that the evolution of the aforementioned systems will support distributed AI both for edge devices and within the mobile network \cite{Net4AI}. Although many researchers argue that the edge devices will not support the use of LLMs, but with continuous breakthroughs in the field of AI, support for edge devices can be extended through distributed learning techniques such as federated learning (FL) and split learning (SL) \cite{PrivateAI}. In addition, quantization and training can be used to fine-tune an LLM on the edge devices. As proposed in \cite{QLoRA}, an LLM with 65 billion parameters can be fine-tuned with quantized low-rank adapters (QLoRA) on a downstream task within a single day, achieving comparable performance compared to other state-of-the-art (SOTA) LLMs. It can be assumed that the convergence of quantized networks, LLMs, 6G, and ZSM-based Multi-Access Edge Computing (MEC) could result in many innovative applications. Researchers have already begun to explore the mutual convergence of LLMs, ZSM, and MEC networks, calling them \enquote{LLMs for networks} and \enquote{networks for LLMs} respectively. We illustrate a network for LLMs that corresponds to the vision of the Network for AI (NetAI) with respect to an integrated 6G-ZSM communication system in Figure \ref{Fig1}. The NetAI vision supports LLM deployments related to the MEC architecture \cite{Net4AI}. Various services such as smart homes, healthcare, education, emergencies, mission-critical applications and finance can be supported with the network for LLMs. \\
\indent Most research today focuses on the integration of LLMs and communication networks, which would undoubtedly bring unprecedented advances and technological innovation. However, one aspect of this technological progress is being overlooked, namely the security aspect. With all the possibilities and potential of LLMs and the ZSM ecosystem, we have to ask ourselves, are LLMs trustworthy? Despite their ability to fine-tune to the downstream task, LLMs are deep neural networks that are vulnerable to privacy attacks, such as model inversion, model poisoning and membership leakage \cite{PrivateAI, nguyen2024}. The growing landscape of LLMs and their integration into communication systems therefore makes it necessary to address security concerns and the development of trustable AI encoders to safeguard the integrity of users and services in ZSM networks. To the best of our knowledge, the studies have not explored the security vulnerabilities in network for LLMs (Net4LLMs), which subsequently leaves us defenseless against such attacks as we progress towards the Net4AI vision.  \\
\indent To address the above problem, in this paper, we propose to audit the trustworthiness of pre-trained AI encoders for membership leakage attacks. The membership leakage attack is an attack in which the adversary tries to find out the distribution of the training data used to train the AI encoders. Considering that the Net4LLMs will focus on fine-tuning the pre-trained AI encoders for service provisioning, the attacker will aim to determine if a data sample was used for the fine-tuning process. The fine-tuning process enables the replacement of task-specific layer in the pre-trained AI encoder to meet tasks such as questions and answers, name entity recognition and classification \cite{nguyen2024, LeakagePI}. We evaluate the trustworthiness of AI encoders against the membership inference attacks. We assume that the adversary has particularly the knowledge of the downstream task and the adversary is provided with the fine-tuned model, also known as black-box setting. We conduct experiments to evaluate the trustworthiness of the AI pre-trained encoders and to develop possible defenses to prevent the membership leakage attacks in the context of Net4LLMs. The specific contributions of this work are characterized as follows:
\begin{itemize}
    \item This is the first study to investigate the trustworthiness of pre-trained AI encoders for Net4LLMs.
    \item Membership leakage attacks in the context of Net4LLMs are explored to assess trustworthiness.
    \item Based on experimental analysis, defenses are proposed to audit trust in Net4LLMs.
    \item At the end of the paper, open issues, challenges and future directions are also proposed to prevent potential adversarial attacks on LLMs. 
\end{itemize}

\begin{figure*}[h]
\centering
  \includegraphics[width=\linewidth]{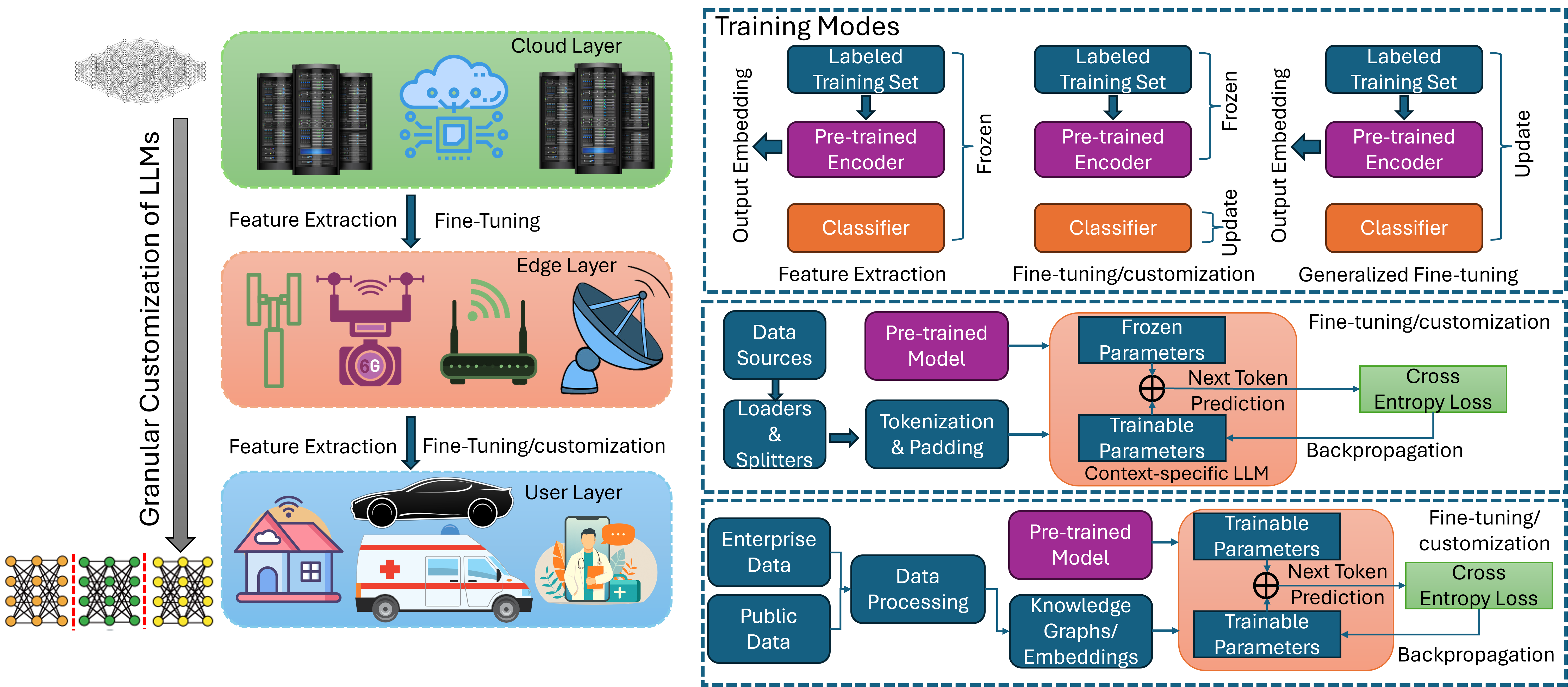}
  \caption{Training, Fine-tuning, and customization strategies for LLMs in ZSM-based MEC framework.}
  \label{Fig2}
\end{figure*}

\section{Related Works}
Recently, we have observed a plethora of advances in the field of LLMs that would be a revolution in the field of communication networks, especially in the design and development of ZSM networks. Some studies have already explored and demonstrated the significance of LLMs for potential ZSM applications. For example, the study in \cite{NetGPT} proposed NetGPT, which enables personalized services to users through generative networks while handling comprehensive network intelligence and cloud collaboration in real time.Xu et al. \cite{LLMmeets6G} focused on the data privacy in 6G communication systems using ZSM and LLMs. The study proposed to design LLM agents based on the principle of split learning by distributing LLMs for different roles across edge devices to make user interaction efficient and collaborative. Their results show that the split learning setting was effective in improving the communication efficiency while offloading the tasks that are complex in nature to the servers for constructing global LLMs. The study in \cite{6Gcomprehensive} emphasized that the newer LLMs must offer multimodal services, i.e. they must handle image, text and audio data in order to offer automated services. Therefore, the deployment of LLM agents in the cloud could pose challenges in terms of data privacy, high bandwidth costs and long response times. However, MEC based on ZSM networks can address the above problems in an effective way. Lin et al. \cite{pushingLLM} also proposed a split learning framework for the deployment of LLMs in 6G-based ZSM networks. However, their work focused on the efficiency and effectiveness of LLMs in terms of parameter sharing, quantization, and efficient fine-tuning rather than data or model security. Nguyen et al. \cite{nguyen2024} highlighted the advantages of using LLMs in ZSM networks while exploring the security vulnerabilities from an adversarial point of view. However, the discussion of model security was very abstract and brief, focusing more on the attacks on services. The study also suggested the use of blockchain technology to avoid the security threats associated with LLMs and ZSM networks. To the best of our knowledge, none of the studies investigated a specific model-based attack scenario related to pre-trained AI encoders and ZSM networks. 

\section{Training/Fine-tuning LLMs in ZSM}

The training strategies for fine-tuning LLMs in ZSM networks are shown in Figure \ref{Fig2}. The strategies are presented in accordance with the MEC framework. The cloud layer can leverage the pre-trained AI encoders of LLMs for any of the training modes, but the downstream task would mostly be generalized when passed to the edge and user layers. However, at the edge and user layer, the customization of the LLMs can be done based on the context and resources. As can be seen from the training modes, fine-tuning the pre-trained encoder and classifier requires large amounts of computational resources. The computational resources decrease significantly when moving to the edge and user layer. In this regard, the edge and user layers can at most fine-tune/adapt the LLMs based on their application by keeping the pre-trained encoder frozen and using the final layers to update the parameters. Alternatively, they can simply freeze the entire LLM and extract only the output embeddings that are used as representative features for training deep neural networks or shallow learning methods.\\
An example of this can be found in Figure \ref{Fig1} for an emergency service application where LLM's pre-trained AI encoder can be used and its parameters frozen while a limited amount of labeled data is used to train the classifier or the final layers. This would be beneficial for edge devices to fine-tune the network locally to improve communication efficiency while utilizing computing resources efficiently. As proposed in \cite{pushingLLM}, the customization can be achieved through a split learning strategy, where the cloud wants to fine-tune the LLM for a specific task with respect to the communication system and splits the network into smaller networks that are trained with devices from the edge layer and the user layer. Some specific techniques such as Low Ranking Adaptation (LoRA), Quantized LoRA (QLoRA), Parameter Efficient Fine-Tuning (PEFT), Deep Speed and ZeRO can be used to fine-tune the LLMs. LoRA uses low-rank approximations to fine-tune the LLMs for specific tasks, resulting in reduced financial and computational costs. QLoRA further reduces memory utilization while fine-tuning LLMs with the LoRA technique. PEFT adjusts key parameters and uses the catastrophic forgetting technique to fine-tune LLMs with a small subset of parameters. ZeRO uses memory optimization and data parallelism techniques to fine-tune LLMs, and DeepSpeed uses the ZeRO redundancy optimizer to fine-tune LLMs in a distributed learning fashion. The above techniques can be used extensively at the user, edge and cloud layers to fine-tune LLMs for various purposes. 

\subsection{Security Issues in LLMs and ZSM}

Several studies have now highlighted the security concerns related to the behavior, architecture and design of LLM. The security concerns arise from the complexity of LLMs and the challenges associated with their deployment and training process. In addition, backdoor attacks are possible in LLMs that cannot be overcome with conventional security measures. These backdoor attacks are applicable to LLMs that are fine-tuned in a supervised manner and trained with adversarial learning or reinforcement learning. The different types of attacks in LLMs deployed within ZSM networks are defined below. 
\begin{itemize}
    \item \textit{Adversarial attacks:} These attacks are carried out by manipulating data to affect the performance of the model. Adversarial attacks can generally be divided into backdoor attacks and poisoning attacks. In the former, a trigger is hidden in the model to manipulate the inference behavior, while in the latter, malicious examples are injected into the training process to deceive the model.
    \item \textit{Inversion attacks:} Inversion attacks are performed to reconstruct the data or to extract certain information from the model gradients. Inversion attacks include replicating the model, extracting training data, gradient leakages, feature space and stealing models. 
    \item \textit{Unfair exploitation and bias attack:} This type of attack is related to the training data used to train or fine-tune the LLMs. The attack disproportionately adds data with a particular label to fine-tune or train the network so that the inference perpetuates biases and unintentionally learns to generate misinformation, social inequalities, reinforcement of stereotypes and discrimination in the generation of responses. 
    \item \textit{Instruction tuning attacks:} These attacks aim to overload the system's resources in order to carry out inadvertent actions. Examples of such attacks are Denial of Service (DoS), indirect prompt injection, jailbreaking and the disclosure of guided prompts. 
    \item \textit{Zero-day attacks:} These attacks are usually called sleeper agents because they are embedded with model weights when a particular defense method fails to eliminate them. This type of attack is usually triggered by specific events or phrases. One example of such attacks is data theft. 
    \item \textit{Inference attacks:} Last but not least, inference attacks aim to extract sensitive information from the model, especially in the context of the training data used to fine-tune a model. Examples of such attacks are attribute inference and membership inference attack. In this paper, we focus on the membership inference attack as it can identify specific data used to train or fine-tune the model. Such information can be used to break the trust and confidentiality of the AI model and be used against the user. Other consequences of membership inference attack include breach of confidentiality, unauthorized access, identity theft and violation of privacy. 
\end{itemize}

\section{Trustable AI encoders}

In this section we provide the information about the threat model, the attack scenario, the datasets used for the attack and the implementation details.

\subsection{Threat Model}

Before we define the threat model, we make some assumptions. We assume that the LLM is pre- trained on a large dataset capable of transforming the input (text) into embeddings. Using the pre-trained AI encoder, a downstream task is fine-tuned by a customized dataset for a specific application in ZSM networks using optimization algorithms and a predefined loss function. The fine-tuned model is then able to transform the input into embeddings or classification probabilities accordingly (which differ from the original, pre-trained AI encoder). We define the attacker's purpose in this scenario as a dichotomous classification problem, where the goal is to determine whether the input provided to the pre-trained AI encoder is a member or non-member of the training dataset used for the subsequent task. In general, existing studies assume two dimensions of an attacker's background knowledge when considering membership inference attacks. 

The first dimension assumes a black box attack, which means that the attacker has no prior knowledge of the pre-trained AI encoder architecture, but the attacker has access to the model that has been trained for the downstream task. This is considered the most realistic scenario, as in ZSM, AI is used as a service and the models adapted for the downstream task would be directly available to the public. The second dimension assumes that the attacker has access to a very small subset of the member training data, which can be used to create an auxiliary dataset. The auxiliary dataset can then be used to train the attack model. Studies have shown that such assumptions can be true if one infers the location and makes an educated guess about the service used in a particular area \cite{ZETA}. With the large plethora of diverse data available on the Internet, it is reasonable to assume that the attacker can gather meaningful data to create a shadow model for lodging membership inference attack that corresponds to a real-world environment. Considering the two dimensions, we assume in this study that the attacker has access to the downstream task model and has some knowledge of the application, which is taken into account by the pre-trained AI encoder of the LLM.

\subsection{Attack Scenario}

It is known that the LLM's pre-trained AI encoder can be used for feature extraction, i.e. the LLM's task of transforming the input into embedding vectors. The mapping of inputs to embeddings benefits the fine-tuning of LLMs or training with deep neural networks for a specific task. However, when the LLM is fine-tuned with the new data for a particular downstream task, it tends to memorize the data during the training process. The memorization suggests that the member data will have higher confidence values compared to the non-member data. Therefore, it can be deduced that: (i) Pre-trained AI encoders of LLMs behave differently to the member and non-member data. (ii) The behavior is propagated to the embedding vectors that are learned during fine-tuning for the downstream tasks, so that the memorization of model will be a part of the downstream model available to the attacker. We intend to use the above features to categorize whether the data is a member or a non-member of the pre-trained AI encoder. We then apply the following steps to evaluate the effectiveness of the attack for a pre-trained AI encoder. 
\begin{itemize}
    \item Our assumptions are that the attacker has some prior knowledge of the application for which the LLM is fine- tuned. Therefore, the attacker scraps the Internet or uses publicly available datasets to create an auxiliary dataset.
    \item The attacker then prepares the auxiliary dataset for training by assigning pesudolabels to the data as members and non-members. The attacker then feeds the pseudo-labeled auxiliary dataset into the downstream model. The training process is then performed to create an attack model that is capable of binary classification, i.e. categorizing the data into members and non-members.
    \item Once the attack model is trained, the attacker can enter the candidate text into the attack model to determine whether the candidate text is a member or a non-member.
\end{itemize}

\subsection{Dataset}

In this work, we use two state-of-the-art pre-trained language models RoBERTa \cite{RoBERTa} and ALBERT \cite{ALBERT} for our experiments. The two language models differ in their training schemes, loss functions and architectures. It should be noted that we have not trained these language models from scratch, but that we use the pre-trained language models for the attack scenario that are publicly available online\footnote{https://huggingface.co/models}. According to the assumption considered for the attack scenario, the attacker has access to the fine-tuned model for the downstream task. Therefore, we consider two publicly available datasets, i.e., Yelp Review/AG's News/SST \cite{Yelp} and CoNLL2003 \cite{CoNLL}. The first data set is intended for the task of text classification, while the second takes into account the task of Named Entity Recognition (NER). To perform the membership inference attack, we use a small portion of the Yelp Review/AG's News/SST and CoNLL2003 dataset, i.e., 0.15\%, of each dataset to construct the auxiliary dataset and label it as member data. We also consider other third-party datasets such as AX, CoLA and IMDB for the auxiliary dataset and label them as non-member data. 

\subsection{Implementation Details}

To perform membership inference attack, we design a five-layer multilayer perceptron as an attack model that uses the output of the model fine-tuned to the downstream task as input. The dimensions of the first layer vary depending on the model fine-tuned to a specific downstream task. Recall, precision and F1 score are used as evaluation metrics for the performance of the attack. The attack model is trained using the ADAM optimizer with a learning rate of $1e-5$. The model is trained for 100 epochs. The auxiliary dataset was divided into two sets, i.e. a test dataset and a training dataset in a ratio of 1:5. 

\begin{figure}[h]
\centering
  \includegraphics[width=\linewidth]{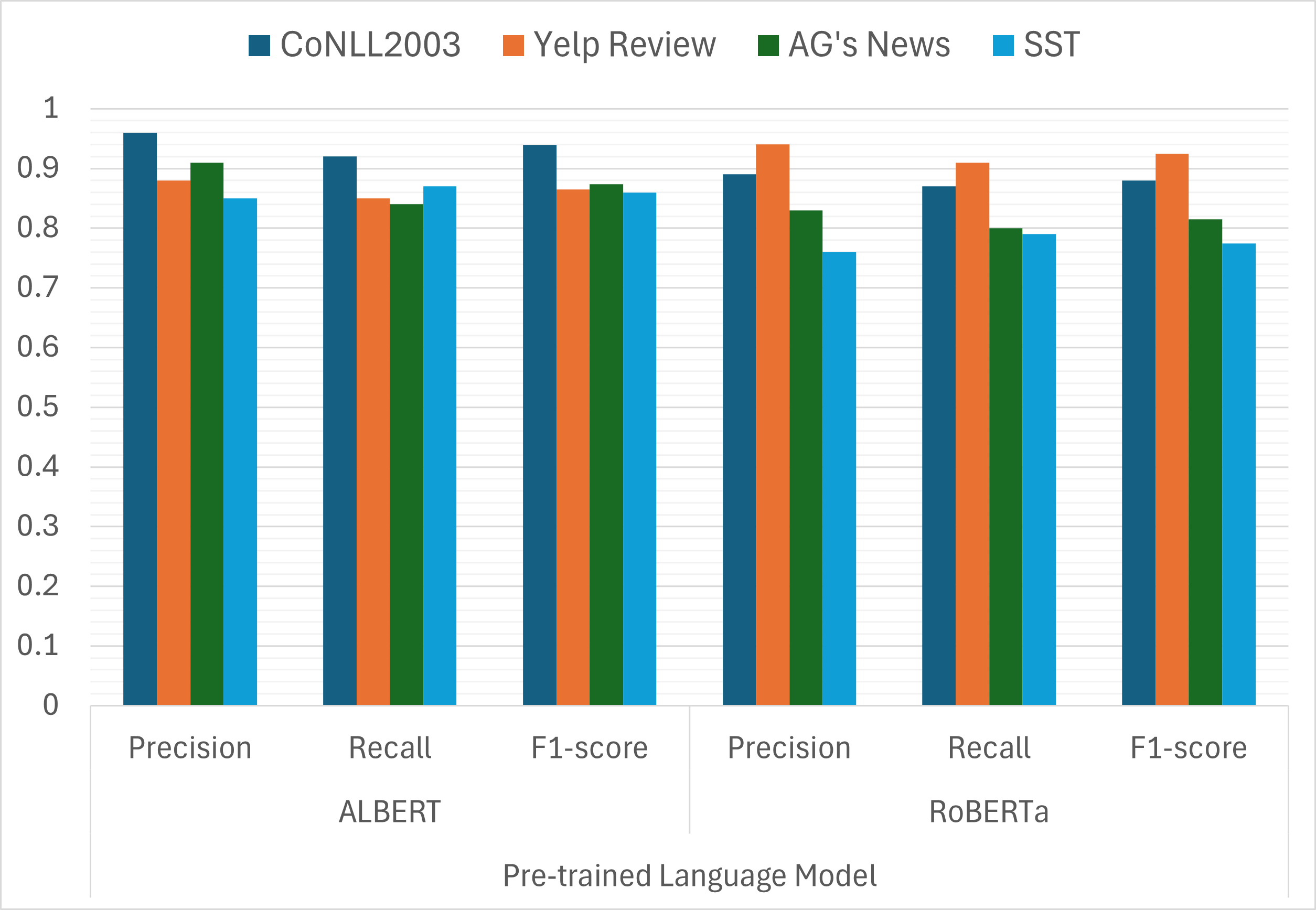}
  \caption{Membership Inference Attack performance on CoNLL2003, Yelp, AG's News and SST downstream tasks. }
  \label{Fig3}
\end{figure}
\begin{figure}[h]
\centering
  \includegraphics[width=\linewidth]{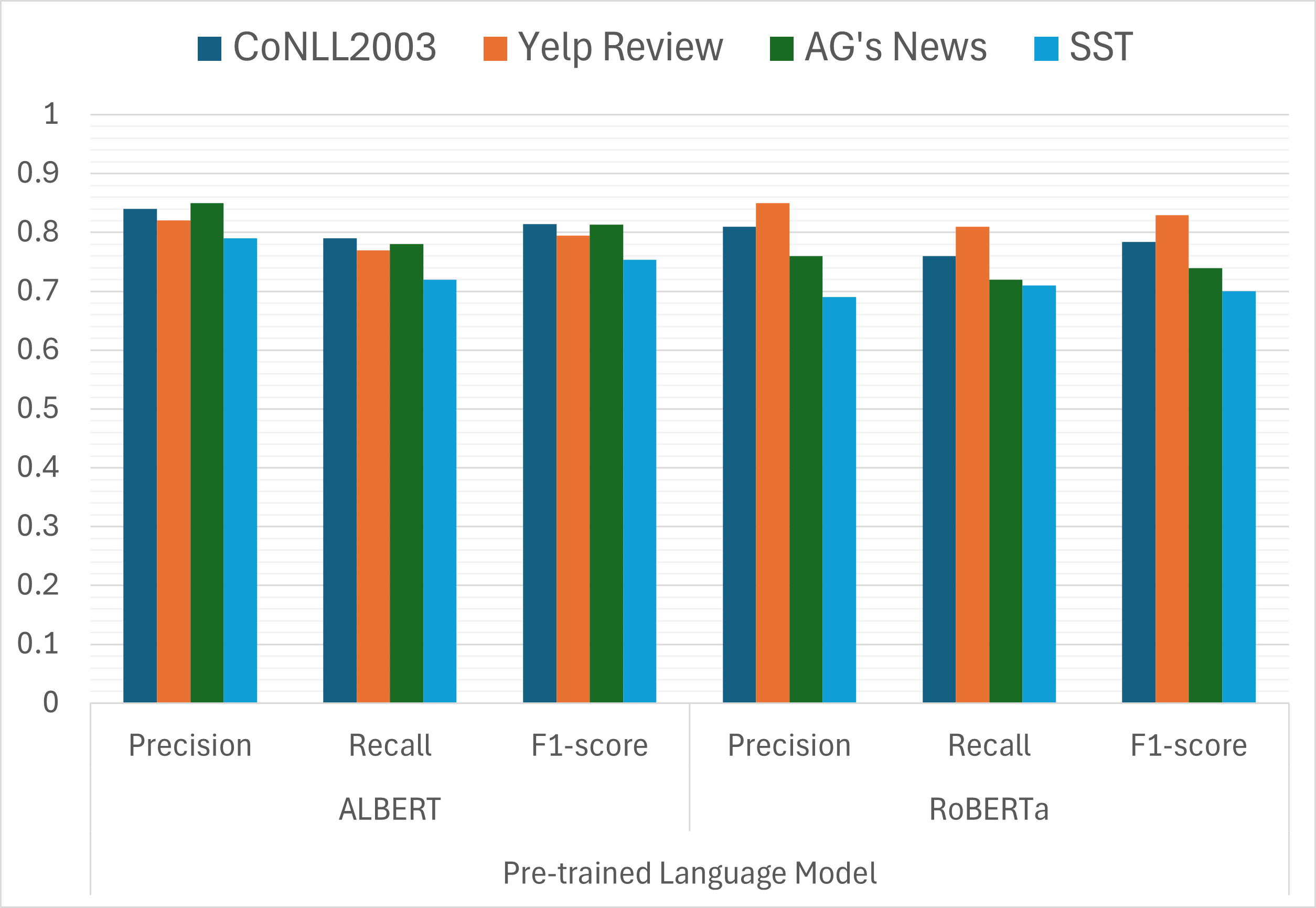}
  \caption{Membership Inference Attack performance on CoNLL2003, Yelp, AG's News and SST downstream tasks when the attack model employs only either of the dataset. }
  \label{Fig4}
\end{figure}
\begin{figure}[h]
\centering
  \includegraphics[width=\linewidth]{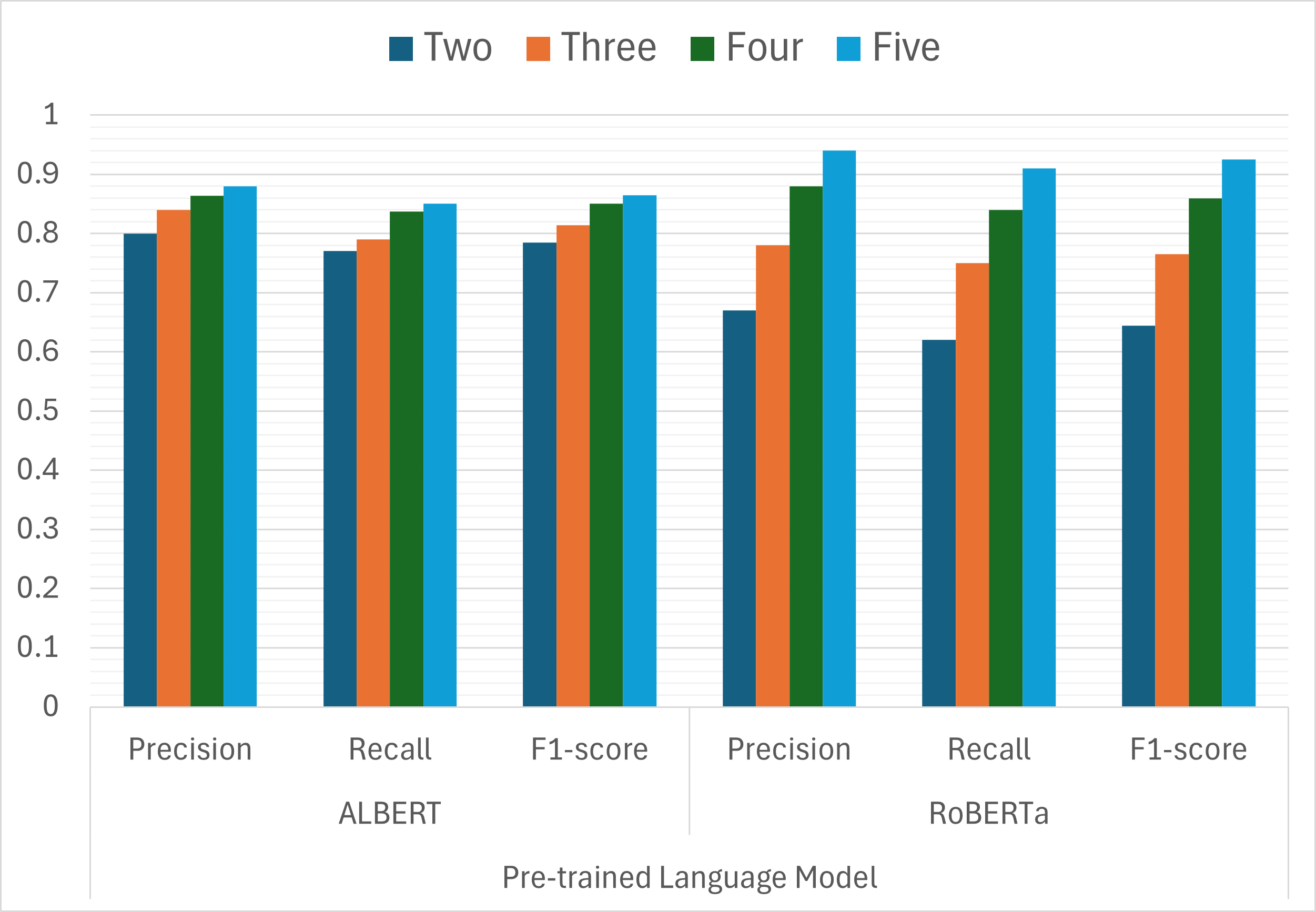}
  \caption{Membership Inference Attack performance on Yelp downstream tasks when varying number of classes. }
  \label{Fig5}
\end{figure}

\section{Experimental analysis}

In Figure \ref{Fig3}, we show the performance of the membership inference attack. It should be noted that Yelp Review/AG's News/SST are classification tasks with 5/4/2 classes, respectively. The baseline, random guessing, refers to the value of $0.5$ for precision and recall. The attack performance shows that the two pre-trained language models achieve a minimum F1 score of $0.77$ in the SST task and a maximum F1 score of 0.94 in the NER task. The results are significantly higher than random guessing, indicating that the membership leak exists in the pre-trained AI encoders. 

The success rate of the attack raises serious concerns about the trustworthiness of LLMs and pre-trained models in ZSM networks. We repeat the above experiment with a reduced auxiliary dataset, i.e., we use Yelp Review/AG's News/SST for training the attack model without considering CoNLL2003 and vice versa to observe the results.  The results for this experiment are shown in Figure 4. It can be seen that the performance is still above the random guess. Furthermore, the performance degradation is about 0.12/0.096 for CoNNL2003, about 0.07/0.095 for Yelp Review, about 0.06/0.075 for AG's News and about 0.105/0.074 for SST, using ALBERT and RoBERTa, respectively. The attack performance still reaches a maximum of 0.83 F1 score if the attacker only has access to the fine-tuned model for the downstream task, but does not use the part of the same datasets. This behavior confirms our assumption that the pre-trained language models memorize the data and behave differently with member and non-member data.

Another interesting aspect was highlighted when we examined the attack performance while varying the number of classes. Since the Yelp dataset has the highest number of classes out of the datasets we selected, namely 5, we varied the number of classes to observe the attack performance. The results of this experiment are shown in Figure 5. It can be seen that the attack performance increases as the number of classes increases. This is very interesting because it shows that the membership inference attack can extract more information from data with a higher number of categories. It also shows why the attack performance on Yelp was better than on AG's News and SST datasets, accordingly. 

\section{Enabling Trust with Pre-trained AI encoders}

Membership leakage and membership inference attacks have been extensively studied in the context of computer vision and image modality. Defenses against such attacks therefore include adversary regularization, differential privacy, data augmentation, adding noise to images, intentional attacks, encryption techniques, and others \cite{PrivateAI,ZETA}. Some of the above techniques are difficult to perform in textual modality, such as addition of noise and intentional attack initialization. Such actions can also degrade the performance of LLMs in ZSM networks. Adding noise to either the data or confidence scores for classification can be used as a defense mechanism. However, studies suggest that such techniques degrade the performance of the downstream task \cite{nguyen2024}.\\
Based on the observations gathered from our experiments, we propose two possible defenses. The first is to reduce the size of the dataset or reduce the number of epochs to fine-tune the network. The intuition is that if the size of the dataset or the number of epochs for fine-tuning in the downstream task is increased, the pre-trained AI encoder would tend to memorize the downstream task data and thus make the membership attack stronger. In this context, we suggest using either active learning or curriculum learning, which can perform the training with less data or fewer epochs.\\
The second defense is based on the intuition \enquote{confidence is defined by trust} (a quote from Patrick Mosher). In this case, however, we would look at confidence and trust from the perspective of AI. We propose to use a trust evaluation module at the edge layer that could evaluate the trust of the pre-trained AI encoder or a fine-tuned LLM with a predefined metric. One of the examples of such a trust evaluation is as follows, assuming the fine-tuned LLM is trained for medical emergency services using a ZSM network.
\begin{itemize}
    \item The LLM is fine-tuned on less number of epochs and a smaller amount of data. The responses do not ask for age or personal information. Results in 88\% for the performance metric.
    \item The LLM is fine-tuned on a large amount of data and high number of epochs. The responses do not ask for age or personal information. Results in 89\% for the performance metric.
    \item The LLM is fine-tuned on a large amount of data and high number of epochs. The responses asks for personal information. Yields 92\% on performance metric. 
\end{itemize}
Given the scenario described above, a trust module might favor the first model as it is less vulnerable to membership leakage attack. The idea is simply to prioritize a fine-tuned LLM for ZSM network based on trustworthiness and confidence scores.

\section{Open Issues, Challenges and Future Directions}

The integration of ZSM and LLMs can be seen as task-oriented communication services, where integration is achieved by utilizing resources from the communication infrastructure, the edge and mobile devices. In return, users receive LLM agents that can perform certain actions, generate data or call application programming interface (API) functions. As already indicated, such integration can lead to security vulnerabilities, including the theft of personal information and more. We have emphasised the importance of a trust module for considered integration, but designing such a trust module can present some challenges. These challenges include, but are not limited to, the following.

\textbf{(i) Active Learning and Curriculum Learning approaches:} One of the ways to cope with membership leakage or inference attack is to use less amount of data and number of epochs for fine-tuning. In this regard, two approaches can be opted for making this possible. The design of trust module needs to favor the model with aforementioned characteristics, however, design of such methods can be a challenge that could help in resisting the attack while not compromising on the performance. 

\textbf{(ii) Multimodal LLMs:} In this paper, we have focused on the LLMs that only work with text data. In reality, users opt for multimodal LLMs that can generate text, images and audio. Each modality has its own security issues when it comes to pre-trained AI encoders. However, it would be quite a challenge to design a trust module for ZSM and LLMs that is suitable for different data modalities.

\textbf{(iii) User Privacy:} Similar to the design of trust modules for different data modalities, training processes, architectural modifications, and encryption techniques must be used to improve user privacy. Various attacks such as model inversion, model poisoning, gradient leakages, and adversarial attacks can be used by attackers to disrupt the services and steal users' private information in ZSM networks. Therefore, in addition to data modality, the trust module must also defend against various privacy attacks.
 
\textbf{(iv) Latency and Bandwidth Issues:} We focused primarily on the trustworthiness of the pre-trained AI encoders. However, the fine-tuning of the LLMs and the deployment of the trust module could burden the services in terms of latency and increased bandwidth. In this context, the selection of suitable models for specific applications, scenarios and needs as well as optimization during deployment must also be researched.

\textbf{(v) Responsible AI:} Finally, this study explores the trust module in the context of security and privacy. The trust module can be explored in the context of responsible use of AI so that hallucinations in the generation of data modality could be controlled or restricted to prevent the spread of misinformation, identity-related attacks and impersonation.

\section{Conclusion}
This article explores the use of LLM deployment in accordance with the ZSM MEC framework. We evaluate the trustworthiness of fine-tuned models to be deployed as services in ZSM networks. We discuss in detail about membership leakage and inference attacks with respect to the textual modality and show that the attacks are quite effective in violating user privacy. We propose possible defense mechanisms to cope with the membership inference attacks and give several open issues, challenges, and research directions for the development of a generalization trust module for LLM deployment in the ZSM-based MEC framework. We believe that this paper provides researchers with the foundation for developing trustworthy LLMs for services in ZSM networks.


\bibliography{ref.bib}

\begin{thebibliography}{10}
\providecommand{\url}[1]{#1}
\csname url@samestyle\endcsname
\providecommand{\newblock}{\relax}
\providecommand{\bibinfo}[2]{#2}
\providecommand{\BIBentrySTDinterwordspacing}{\spaceskip=0pt\relax}
\providecommand{\BIBentryALTinterwordstretchfactor}{4}
\providecommand{\BIBentryALTinterwordspacing}{\spaceskip=\fontdimen2\font plus
\BIBentryALTinterwordstretchfactor\fontdimen3\font minus \fontdimen4\font\relax}
\providecommand{\BIBforeignlanguage}[2]{{%
\expandafter\ifx\csname l@#1\endcsname\relax
\typeout{** WARNING: IEEEtran.bst: No hyphenation pattern has been}%
\typeout{** loaded for the language `#1'. Using the pattern for}%
\typeout{** the default language instead.}%
\else
\language=\csname l@#1\endcsname
\fi
#2}}
\providecommand{\BIBdecl}{\relax}
\BIBdecl

\bibitem{surveyLLM}
L.~Wang, C.~Ma, X.~Feng, Z.~Zhang, H.~Yang, J.~Zhang, Z.~Chen, J.~Tang, X.~Chen, Y.~Lin, W.~X. Zhao, Z.~Wei, and J.~Wen, ``{A Survey on Language Model based Autonomous Agents},'' \emph{Frontiers of Computer Science}, vol.~18, no.~6, p. 186345, 2024.

\bibitem{Net4AI}
W.~Tong and P.~Zhu, ``{6G: The Next Horizon From Connected People and Things to Connected Intelligence},'' Huawei Technology, Tech. Rep., 2022.

\bibitem{PrivateAI}
S.~A. Khowaja, K.~Dev, N.~M.~F. Qureshi, P.~Khuwaja, and L.~Foschini, ``Toward industrial private ai: A two-tier framework for data and model security,'' \emph{IEEE Wireless Communications}, vol.~29, no.~2, pp. 76--83, 2022.

\bibitem{QLoRA}
T.~Dettmers, A.~Pagnoni, A.~Holtzman, and L.~Zettlemoyer, ``{QLoRA: Efficient Finetuning of Quantized LLMs},'' in \emph{Advances in Neural Information Processing Systems}, vol.~36, 2023, pp. 10\,088--10\,115.

\bibitem{nguyen2024}
\BIBentryALTinterwordspacing
T.~Nguyen, H.~Nguyen, A.~Ijaz, S.~Sheikhi, A.~V. Vasilakos, and P.~Kostakos, ``Large language models in 6g security: challenges and opportunities,'' 2024. [Online]. Available: \url{https://arxiv.org/abs/2403.12239}
\BIBentrySTDinterwordspacing

\bibitem{LeakagePI}
N.~Lukas, A.~Salem, R.~Sim, S.~Tople, L.~Wutschitz, and S.~Zanella-Béguelin, ``Analyzing leakage of personally identifiable information in language models,'' in \emph{2023 IEEE Symposium on Security and Privacy (SP)}, 2023, pp. 346--363.

\bibitem{NetGPT}
Y.~Chen, R.~Li, Z.~Zhao, C.~Peng, J.~Wu, E.~Hossain, and H.~Zhang, ``Netgpt: An ai-native network architecture for provisioning beyond personalized generative services,'' \emph{IEEE Network}, vol. Early Access, pp. 1--10, 2024.

\bibitem{LLMmeets6G}
\BIBentryALTinterwordspacing
M.~Xu, D.~Niyato, J.~Kang, Z.~Xiong, S.~Mao, Z.~Han, D.~I. Kim, and K.~B. Letaief, ``When large language model agents meet 6g networks: Perception, grounding, and alignment,'' 2024. [Online]. Available: \url{https://arxiv.org/abs/2401.07764}
\BIBentrySTDinterwordspacing

\bibitem{6Gcomprehensive}
\BIBentryALTinterwordspacing
S.~Long, F.~Tang, Y.~Li, T.~Tan, Z.~Jin, M.~Zhao, and N.~Kato, ``6g comprehensive intelligence: network operations and optimization based on large language models,'' 2024. [Online]. Available: \url{https://arxiv.org/abs/2404.18373}
\BIBentrySTDinterwordspacing

\bibitem{pushingLLM}
\BIBentryALTinterwordspacing
Z.~Lin, G.~Qu, Q.~Chen, X.~Chen, Z.~Chen, and K.~Huang, ``Pushing large language models to the 6g edge: Vision, challenges, and opportunities,'' 2024. [Online]. Available: \url{https://arxiv.org/abs/2309.16739}
\BIBentrySTDinterwordspacing

\bibitem{ZETA}
S.~A. Khowaja, P.~Khuwaja, K.~Dev, K.~Singh, L.~Nkenyereye, and D.~Kilper, ``Zeta: Zero-trust attack framework with split learning for autonomous vehicles in 6g networks,'' in \emph{2024 IEEE Wireless Communications and Networking Conference (WCNC)}, 2024, pp. 1--6.

\bibitem{RoBERTa}
\BIBentryALTinterwordspacing
Y.~Liu, M.~Ott, N.~Goyal, J.~Du, M.~Joshi, D.~Chen, O.~Levy, M.~Lewis, L.~Zettlemoyer, and V.~Stoyanov, ``Roberta: A robustly optimized bert pretraining approach,'' 2019. [Online]. Available: \url{https://arxiv.org/abs/1907.11692}
\BIBentrySTDinterwordspacing

\bibitem{ALBERT}
\BIBentryALTinterwordspacing
Z.~Lan, M.~Chen, S.~Goodman, K.~Gimpel, P.~Sharma, and R.~Soricut, ``Albert: A lite bert for self-supervised learning of language representations,'' in \emph{International Conference on Learning Representations}, 2020. [Online]. Available: \url{https://openreview.net/forum?id=H1eA7AEtvS}
\BIBentrySTDinterwordspacing

\bibitem{Yelp}
\BIBentryALTinterwordspacing
X.~Zhang, J.~Zhao, and Y.~LeCun, ``Character-level convolutional networks for text classification,'' in \emph{Advances in Neural Information Processing Systems}, vol.~28.\hskip 1em plus 0.5em minus 0.4em\relax Curran Associates, Inc., 2015, pp. 1--9. [Online]. Available: \url{https://proceedings.neurips.cc/paper_files/paper/2015/file/250cf8b51c773f3f8dc8b4be867a9a02-Paper.pdf}
\BIBentrySTDinterwordspacing

\bibitem{CoNLL}
\BIBentryALTinterwordspacing
X.~Li, J.~Feng, Y.~Meng, Q.~Han, F.~Wu, and J.~Li, ``A unified {MRC} framework for named entity recognition,'' in \emph{Proceedings of the 58th Annual Meeting of the Association for Computational Linguistics}.\hskip 1em plus 0.5em minus 0.4em\relax Association for Computational Linguistics, Jul. 2020, pp. 5849--5859. [Online]. Available: \url{https://aclanthology.org/2020.acl-main.519}
\BIBentrySTDinterwordspacing

\end{thebibliography}
\bibliographystyle{IEEEtran}
\end{document}